\documentstyle[epsfig]{aipproc}

\def\beq{\begin{equation}}
\def\eeq{\end{equation}}
\def\bea{\begin{eqnarray}}
\def\eea{\end{eqnarray}}
\def\bq{\begin{quote}}
\def\eq{\end{quote}}

\def\NP{{\it Nucl.Phys.} }
\def\PL{{\it Phys.Lett.} }
\def\PR{{\it Phys.Rev.} }
\def\PRL{{\it Phys.Rev.Lett.} }

\def\gappeq{\mathrel{\rlap {\raise.5ex\hbox{$>$}}
{\lower.5ex\hbox{$\sim$}}}}

\def\lappeq{\mathrel{\rlap{\raise.5ex\hbox{$<$}}
{\lower.5ex\hbox{$\sim$}}}}

\begin{document}
\begin{flushright}
CERN-TH/99-41 \\
astro-ph/9903003
\end{flushright}
\title{Limits on Sparticle Dark Matter}

\author{John Ellis$^*$ }
\address{$^*$Theoretical Physics Division, CERN \\ CH -- 1211 Geneva 23}
\maketitle

\begin{center}
{\it Talk presented at COSMO 98, Asilomar, California, November 1998}
\end{center}

\begin{abstract}
Arguments are given that the  lightest supersymmetric
particle should be a neutralino $\chi$. Minimizing the fine
tuning
of the gauge hierarchy favours $\Omega_{\chi} h^2 \sim 0.1$.
There are important constraints on the parameter space os the MSSM from the stability of
the electroweak vacuum. Co-annihilation with the next-to-lightest
supersymmetric particle is potentially significant. Incorporating the latest
accelerator constraints from LEP and elsewhere, we find that 50 GeV $\lappeq
m_\chi \lappeq$ 600 GeV and $\tan\beta \gappeq$ 2.5, if soft supersymmetry
breaking parameters are assumed to be universal.
\end{abstract}

\section*{The Lightest Supersymmetric Particle in the MSSM}
 
 The motivation for supersymmetry at an accessible energy is provided by 
the gauge
 hierarchy problem~\cite{hierarchy}, namely that of understanding why $m_W
\ll m_P$, the 
only
 candidate for a fundamental mass scale in physics. Alternatively and
 equivalently, one may ask why $G_F\sim g^2/m^2_W \gg G_N = 1/m^2_P$,
where $M_P$ is the Planck mass, expected to be the fundamental
gravitational mass scale. Or 
one may
 ask why
 the Coulomb potential inside an atom is so much
 larger than the Newton potential, which is equivalent to why $e^2 = 
{\cal O}(1) \gg
 m_pm_e/m^2_P$, where $m_{p,e}$ are the proton and electron masses.

One might think it would be sufficient to choose the bare mass
 parameters: $m_W\ll m_P$. However, one must then contend with quantum
 corrections, which are quadratically divergent:
 \beq
 \delta m^2_{H,W} = {\cal O}~\left({\alpha\over\pi}\right)~\Lambda^2
 \label{twentysix}
 \eeq
 which is much larger than $m_W$, if the cutoff $\Lambda$ representing 
the
 appearance of new physics is taken to be ${\cal O}(m_P)$. This means 
that one must
 fine-tune the bare mass parameter so that it is almost exactly cancelled 
by the
 quantum correction (\ref{twentysix}) in order to obtain a small physical 
value of
 $m_W$. This seems unnatural, and the alternative is to introduce new 
physics at
 the TeV scale, so that the correction (\ref{twentysix}) is naturally 
small.
 
 At one stage, it was proposed that this new physics might correspond to 
the Higgs
 boson being composite~\cite{technicolour}. However, calculable scenarios
of this type are
 inconsistent with the precision electroweak data from LEP and elsewhere. 
The
 alternative is to postulate approximate supersymmetry~\cite{susy}, whose
pairs of 
bosons and
 fermions produce naturally cancelling quantum corrections:
 \beq
 \delta m^2_W = {\cal O}~\left({\alpha\over\pi}\right)~\vert m^2_B - 
m^2_F\vert
 \label{twentyseven}
 \eeq
 that are naturally small: 
 $\delta m^2_W \lappeq m^2_W$ if 
\beq
\vert m^2_B - m^2_F\vert \lappeq {\rm 1 TeV}^2.
\label{twentysevenhalf}
\eeq
 There are many other possible motivations for supersymmetry, but this is 
the only
 one that gives reason to expect that it might be accessible to the 
current
 generation of accelerators and in the range  expected for a 
cold
 dark matter particle.
 
 The minimal supersymmetric extension of the Standard Model (MSSM) has 
the same
 gauge interactions as the Standard Model, and the Yukawa interactions 
are very
 similar:
 \beq
 \lambda_d QD^cH + \lambda_\ell LE^cH + \lambda_u QU^c\bar H + \mu\bar HH
 \label{twentyeight}
 \eeq
 where the capital letters denote supermultiplets with the same quantum 
numbers as
 the left-handed fermions of the Standard Model. The couplings
 $\lambda_{d,\ell,u}$ give masses to down quarks, leptons and up quarks
 respectively, via distinct Higgs fields $H$ and $\bar H$, which are 
required in
 order to cancel triangle anomalies. The new parameter in 
(\ref{twentyeight}) is the
 bilinear coupling $\mu$ between these Higgs fields, that plays a 
significant
 r\^ole in the description of the lightest supersymmetric particle, as we 
see
 below. The gauge quantum numbers do not forbid the appearance of 
additional
 couplings
 \beq
 \lambda LLE^c + \lambda^\prime LQD^c + \lambda U^cD^cD^c
 \label{twentynine}
 \eeq
 but these violate lepton or baryon number, and we assume they are 
absent.
 One significant aspect of the MSSM is that the quartic scalar 
interactions are
 determined, leading to important constraints on the Higgs mass, as we 
also see
 below.
 
 Supersymmetry must be broken, since  supersymmetric partner particles do 
not have
 identical masses, and this is usually parametrized by scalar mass 
parameters
 $m^2_{0_i}\vert\phi_i\vert^2$, gaugino masses ${1\over 2} M_a\tilde
 V_a\cdot\tilde V_a$ and trilinear scalar couplings $A_{ijk}\lambda_{ijk}
 \phi_i\phi_j\phi_k$. These are commonly supposed to be inputs from some
 high-energy physics such as supergravity or string theory. It is often
 hypothesized that these inputs are universal: $m_{0_i} \equiv m_0, 
M_a\equiv
 M_{1/2}, A_{ijk}\equiv A$, but these assumptions are not strongly 
motivated by any
 fundamental theory. The physical sparticle mass parameters are then 
renormalized  in a calculable way:
 \beq
 m^2_{0_i} = m^2_0 + C_i m^2_{1/2}~,~~ M_a = \left({\alpha_a\over
 \alpha_{GUT}}\right)~~m_{1/2}
 \label{thirty}
 \eeq
where the $C_i$ are calculable coefficients~\cite{renorm}
 and MSSM phenomenology is then parametrized by $\mu, m_0, m_{1/2}, A$ 
and
 $\tan\beta$ (the ratio of Higgs v.e.v.'s).
 
 Precision electroweak data from LEP and elsewhere provide two 
qualitative
 indications in favour of supersymmetry. One is that the inferred 
magnitude of
 quantum corrections favour a relatively light Higgs boson~\cite{LEPEWWG}
 \beq
 m_h = 66^{+74}_{-39} \pm 10~{\rm GeV}
 \label{thirtyone}
 \eeq
 which is highly consistent with the value predicted in the MSSM: $m_h 
\lappeq$
 150 GeV~\cite{susymh} as a result of the constrained quartic couplings.
(On the other 
hand,
 composite Higgs models predicted an effective Higgs mass $\gappeq$ 1 TeV
and other unseen quantum corrections.)  The other indication in favour 
of
 low-energy supersymmetry is provided by measurements of the gauge 
couplings at
 LEP, that correspond to $\sin^2 \theta_W \simeq 0.231$ in agreement with 
the
 predictions of supersymmetric GUTs with sparticles weighing about 1~TeV, 
but
 in disagreement with non-supersymmetric GUTs that predict
 $\sin^2\theta_W \sim 0.21$ to 0.22~\cite{sintheta}.  Neither of these
arguments provides 
an accurate
estimate of the sparticle mass scales, however, since they are both only 
logarithmically
sensitive to $m_0$ and/or $m_{1/2}$.

The lightest supersymmetric particle (LSP) is expected to be stable in 
the MSSM, and hence
should be present in the Universe today as a cosmological relic from the 
Big Bang~\cite{EHNOS}.  This is
a consequence of a multiplicatively-conserved quantum number called $R$ 
parity, which is
related to baryon number, lepton number and spin:
\beq
R = (-1)^{3B+L+2S}
\label{thirtytwo}
\eeq
It is easy to check that $R = +1$ for all Standard Model particles and $R 
= -1$ for all
their supersymmetric partners.  The interactions (\ref{twentynine}) would 
violate $R$, but
not a Majorana neutrino mass term or the other interactions in $SU(5)$ or 
$SO(10)$ GUTs. 
There are three important consequences of $R$ conservation: (i) 
sparticles are always
produced in pairs, e.g., $pp \to \tilde{q} \tilde{g} X$, $e^+ e^- \to 
\tilde{\mu}^+
\tilde{\mu}^-$,  (ii) heavier sparticles decay into lighter sparticles, 
e.g., $\tilde{q}
\to q \tilde{g}$,
$\tilde{\mu} \to \mu \tilde{\gamma}$, and (iii) the LSP is stable because 
it has no legal
decay mode.

If such a supersymmetric relic particle had either electric charge or 
strong interactions,
it would have condensed along with ordinary baryonic matter during the 
formation of
astrophysical structures, and should be present in the Universe today in 
anomalous heavy
isotopes.  These have not been seen in studies of $H$, $He$, $Be$, $Li$, 
$O$, $C$, $Na$,
$B$ and $F$ isotopes at levels ranging from $10^{-11}$ to
$10^{-29}$~\cite{Smith}, 
which are far below
the calculated relic abundances from the Big Bang:
\beq
\frac{n_{relic}}{n_p} \; \gappeq \; 10^{-6} \; \mbox{to} \; 10^{-10}
\label{thirtythree}
\eeq
 for relics with electromagnetic or strong interactions. Except
possibly for very heavy relics, one would expect these primordial relic
particles to condense into galaxies, stars and planets, along with
ordinary bayonic material, and hence show up as an anaomalous heavy
isotope of one or more of the elements studied. There
would also be a `cosmic rain' of
such relics~\cite{Nussinov}, but this would presumably not be the dominant
source
of such particles on earth. The conflict with
(\ref{thirtythree}) is sufficiently acute that
the lightest supersymmetric
relic must presumably be electromagnetically neutral and weakly
interacting~\cite{EHNOS}.  In particular, I
believe that the possibility of a stable gluino can be excluded.  
This leaves as
scandidates for cold dark matter a sneutrino $\tilde{\nu}$ with spin 0, 
some neutralino
mixture of $\tilde{\gamma} / \tilde{H}^0 / \tilde{Z}$ with spin 1/2, and 
the gravitino
$\tilde{G}$ with spin 3/2.

LEP searches for invisible $Z^0$ decays require $m_{\tilde{\nu}} \, 
\gappeq \, 43 \;
\mbox{GeV}$~\cite{EFOS}, and searches for the interactions of relic
particles with 
nuclei then enforce 
$m_{\tilde{\nu}} \, \gappeq \,$ few  TeV~\cite{Klap}, so we exclude this
possibility 
for the
LSP.  The possibility of a gravitino $\tilde{G}$ LSP has attracted 
renewed interest
recently with the revival of gauge-mediated models of supersymmetry 
breaking~\cite{GR}, and could
constitute warm dark matter if $m_{\tilde{G}} \simeq 1 \, \mbox{keV}$.  
In this talk,
however, I concentrate on the $\tilde{\gamma} / \tilde{H}^0 /
\tilde{Z}^0$ neutralino combination $\chi$, which is the best 
supersymmetric candidate for
cold dark matter.

The neutralinos and charginos may be characterized 
at the tree level by three parameters: 
$m_{1/2}$, $\mu$
and tan$\beta$.  The lightest neutralino $\chi$ simplifies in the limit 
$m_{1/2} \to 0$
where it becomes essentially a pure photino $\tilde{\gamma}$, or $\mu \to 
0$ where it
becomes essentially a pure higgsino $\tilde{H}$.  These possibilities are 
excluded,
however, by LEP and the FNAL Tevatron collider~\cite{EFOS}.  From the
point of view 
of astrophysics and
cosmology, it is encouraging that there are generic domains of the 
remaining parameter
space where $\Omega_{\chi}h^2 \simeq 0.1$ to $1$, in particular in 
regions where $\chi$ is
approximately a $U(1)$ gaugino $\tilde{B}$, as seen in Fig.
1~\cite{EFGOS}.

\begin{figure}[htb] 
\centerline{\epsfig{file=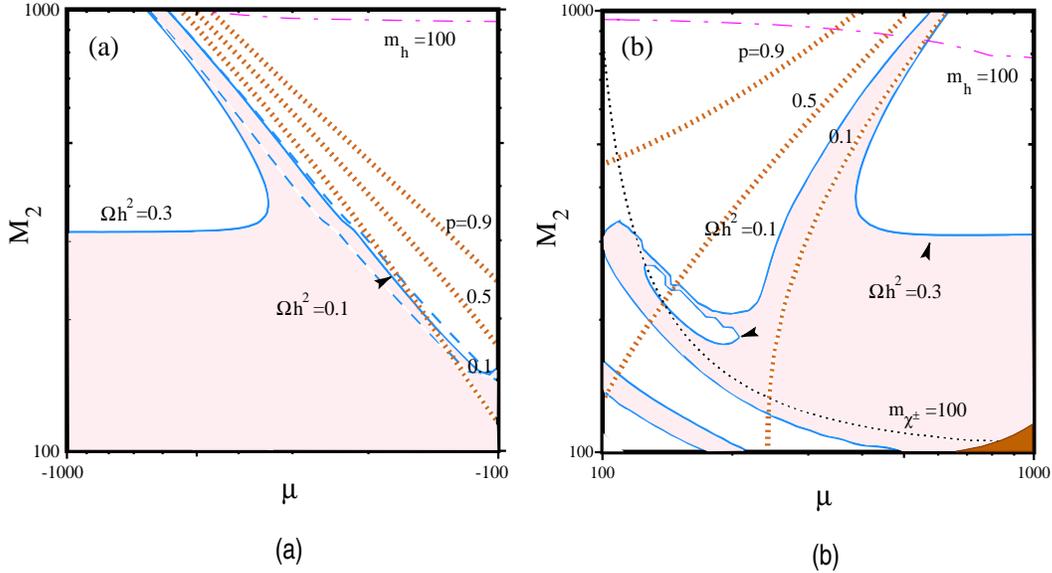,height=3in,width=5.5in}}
\vspace{10pt}
\hspace{3pt}
\caption{Regions of the $(\mu, M_2)$ plane in which the
supersymmetric relic density may lie within the interesting range
$0.1 \le \Omega h^2 \le 0.3$~[14].}
\label{fig1}
\end{figure}

\begin{figure}[htb] 
\centerline{\epsfig{file=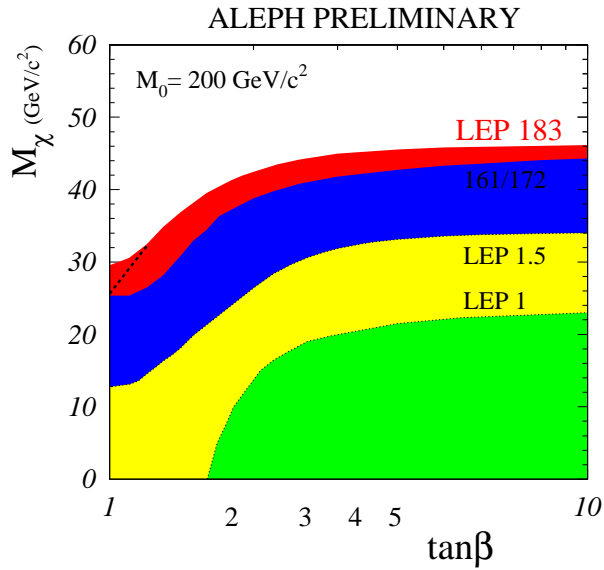,height=3.5in,width=3.5in}}
\vspace{10pt}
\hspace{3pt}
\caption[]{Experimental lower limit on the lightest neutralino mass,
inferred from unsuccessful chargino and neutralino searches at
LEP~\cite{LEPC}.}
\label{fig2}
\end{figure}

\begin{figure}[htb] 
\centerline{\epsfig{file=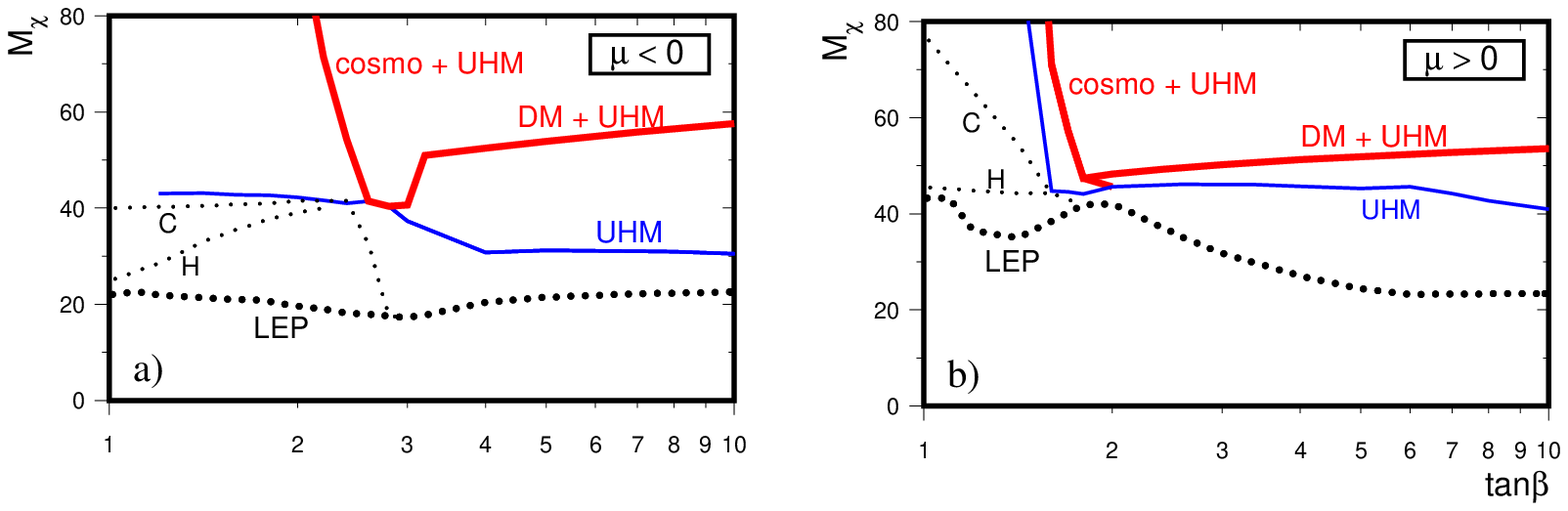,height=2.3in,width=5in}}
\vspace{10pt}
\caption[]{Theoretical lower limits on the lightest neutralino
mass, obtained by using the unsuccessful Higgs searches (H), the
cosmological upper limit on the relic density (C), the assumption that
all input scalar masses are universal, including those of the Higgs
multiplets (UHM), and combining this with the cosmological upper (cosmo)
and astrophysical lower (DM) limits on the cold dark matter
density~\cite{EFOS}.}
\label{fig3}
\end{figure}

Purely experimental searches at LEP enforce $m_{\chi} \gappeq 30$ GeV, as 
seen in Fig. 2~\cite{LEPC}. 
This bound can be strengthened by making various theoretical assumptions, 
such as the
universality of scalar masses $m_{0_i}$, including in the Higgs sector, 
the cosmological
dark matter requirement that $\Omega_{\chi} h^2 \leq 0.3$ and the 
astrophysical preference
that $\Omega_{\chi} h^2 \geq 0.1$.  Taken together as in Fig. 3, we see 
that they enforce
\beq
m_{\chi} \gappeq 42 \; \mbox{GeV}
\label{thirtyfour}
\eeq
and LEP should eventually be able to establish or exclude $m_{\chi}$ up 
to about 50 GeV. 
As seen in Fig. 4, LEP has already explored almost all the parameter 
space available for a
Higgsino-like LSP, and this possibility will also be thoroughly explored 
by LEP~\cite{LEPC}.

\begin{figure}[htb] 
\centerline{\epsfig{file=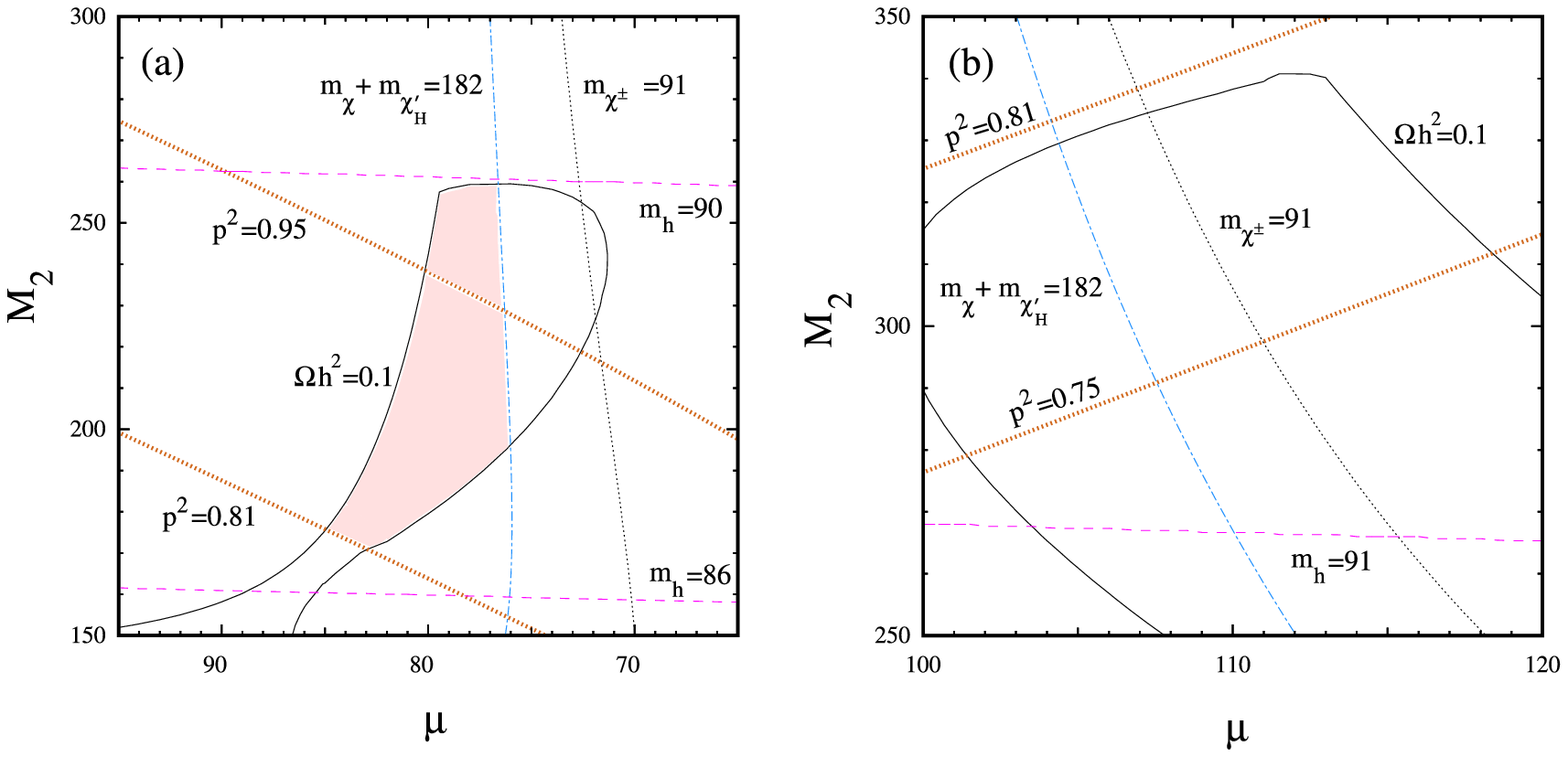,height=2in,width=5in}}
\vspace{10pt}
\caption{The regions of the $(\mu, M_2)$ plane where the
lightest supersymmetric particle may still be a Higgsino, taking
into account the indicated LEP constraints~[14]. The
Higgsino purity is indicated by $p^2$.}
\label{fig4}
\end{figure}

\begin{figure}[htb] 
\centerline{\epsfig{file=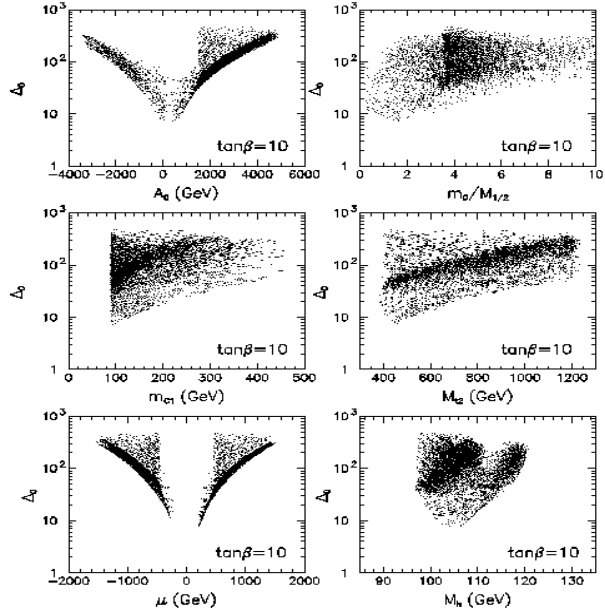,height=3.5in,width=3.5in}}
\vspace{10pt}
\caption{The fine-tuning price $\Delta_0$ imposed by LEP for
tan$\beta = 10$, as a function of model parameters~[17].}
\label{fig5}
\end{figure}

\begin{figure}[htb] 
\centerline{\epsfig{file=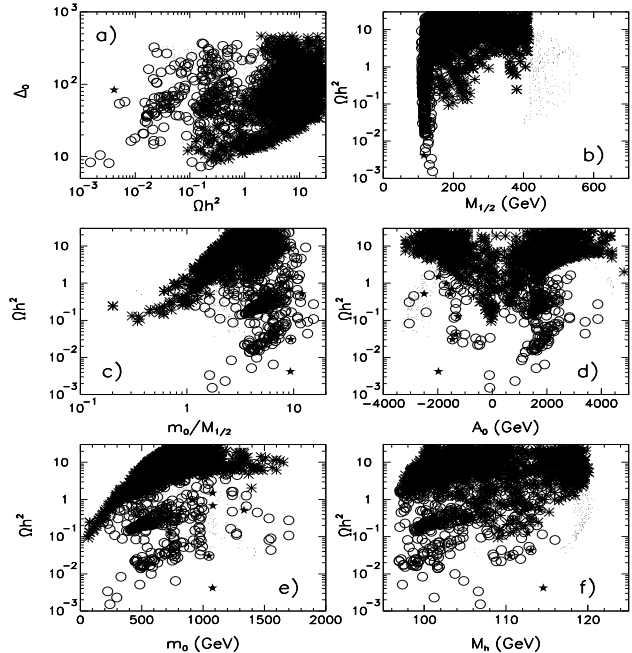,height=3.5in,width=3.5in}}
\vspace{10pt}
\caption{The correlation between the fine-tuning price $\Delta_0$
and the relic density $\Omega h^2$, showing dependences on model
parameters~[18].}
\label{fig6}
\end{figure}

\begin{figure}[htb] 
\centerline{\epsfig{file=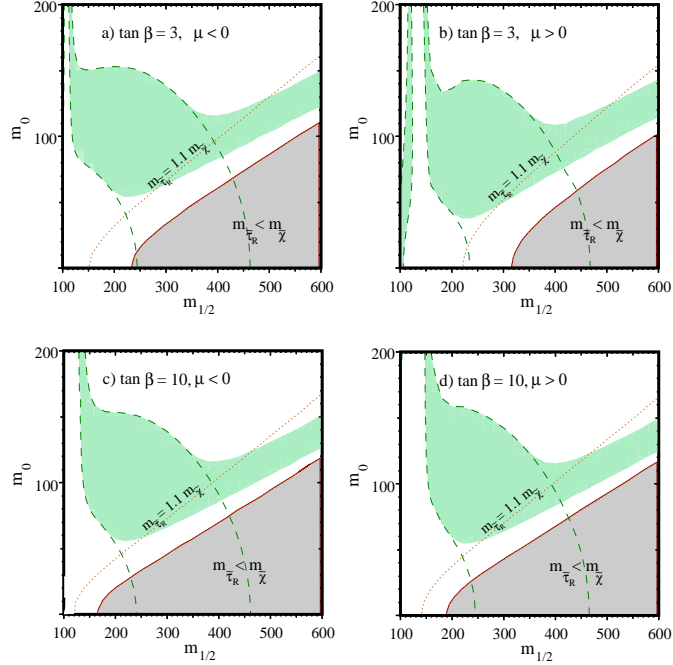,height=3.5in,width=3.5in}}
\vspace{10pt}
\caption{The change in the domain of parameter space
allowed by the requirements $0.1 \le \Omega h^2 \le 0.3$ after 
(shaded region) and before (dashed lines) including $\tilde \tau$
co-annihilation~[20].}
\label{fig7}
\end{figure}

\begin{figure}[htb] 
\centerline{\epsfig{file=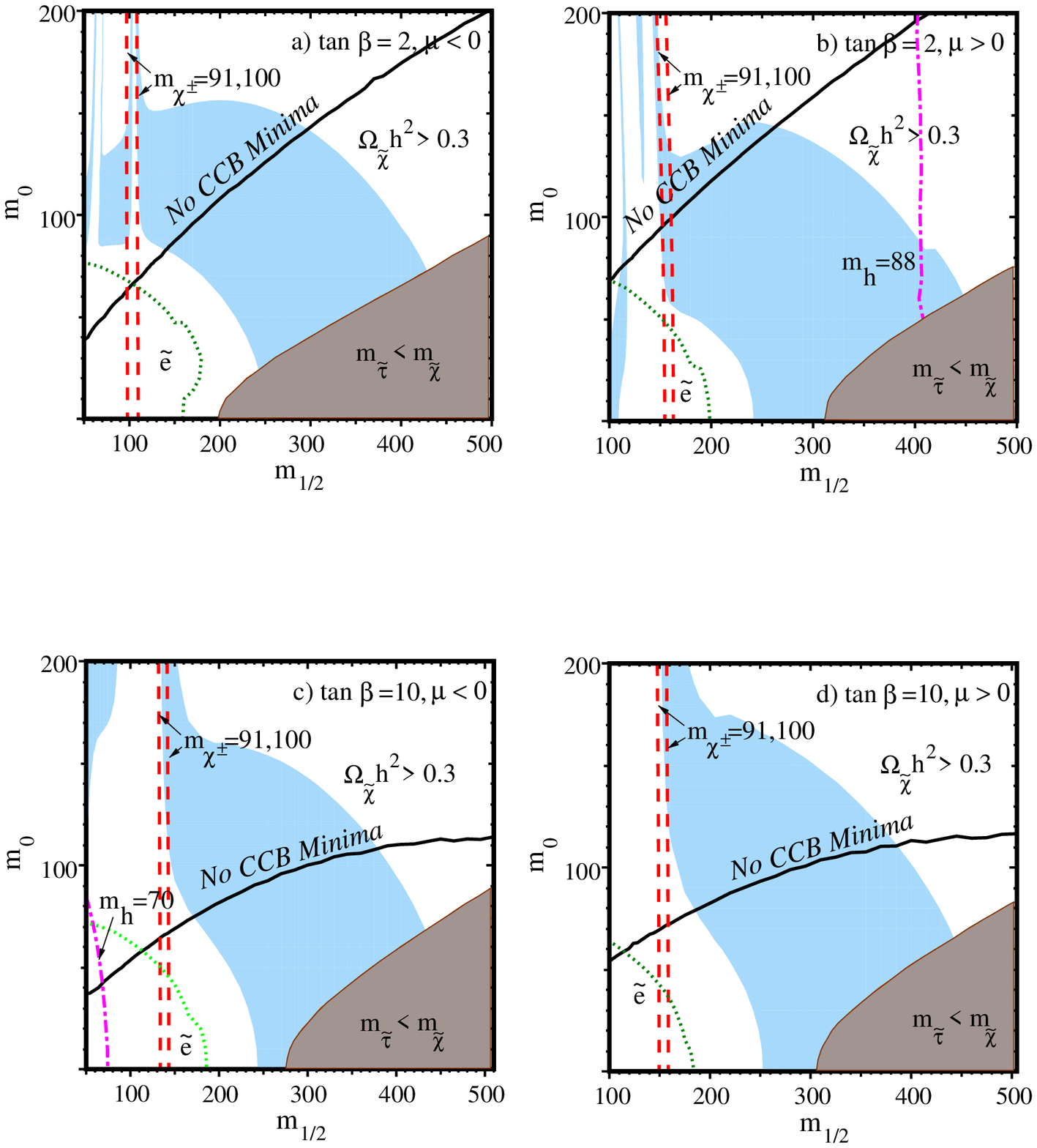,height=3.5in,width=3.5in}}
\vspace{10pt}
\caption{ The combined cosmological and experimental
constraints on the constrained MSSM, for $\tan\beta=2,10$ and both $\mu<0$
and $\mu>0$.  The dashed contours represent current and future LEP
chargino bounds, dotted contours are slepton bounds, and dot-dashed
contours are Higgs bounds.  The light-shaded region gives
$0.1<\Omega_\chi h^2<0.3$.  Below the solid contour, CCB minima are present in
the $LLE, LH_2$ direction.  The value $A_0=-m_{1/2}$ has been chosen,
so as to minimise the area containing CCB minima~[19].}
\label{fig8}
\end{figure}

\section*{What is the ``Natural" Relic LSP Density?}

Should one be concerned that no sparticles have yet been seen by either 
LEP or the FNAL
Tevatron collider?  One way to quantify this is via the amount of 
fine-tuning of the input
parameters required to obtain the physical value of $m_W$~\cite{fine}:
\beq
\Delta_o = 
Max_{i} \;
\mid \frac{a_i}{m_W} \; \frac{\partial m_W}{\partial a_i} \mid
\label{thirtyfive}
\eeq
where $a_i$ is a generic supergravity input parameter.  As seen in Fig. 
5, the LEP
exclusions impose~\cite{CEOP}
\beq
\Delta_o \gappeq 8
\label{thirtysix}
\eeq
Although fine-tuning is a matter of taste, this is perhaps not large 
enough to be alarming,
and could in any case be reduced significantly if a suitable theoretical 
relation between
some input parameters is postulated~\cite{CEOP}.  

It is interesting to
note that the 
amount of
fine-tuning $\Delta_o$ is minimized when $\Omega_{\chi}h^2 \sim 0.1$ as
preferred astrophysically, as seen in Fig. 6~\cite{CEOPO}. This means that
solving the 
gauge hierarchy
problem naturally leads to a relic neutralino density in the range of 
interest to
astrophysics and cosmology.  I am unaware of any analogous argument for 
the neutrino or the
axion.

\section*{Is our Electroweak Vacuum Stable?}
For certain ranges of the MSSM parameters, our present electroweak vacuum is
unstable against the development of vev's for $\tilde q$ and $\tilde l$
fields, leading to vacua that would break charge and colour conservation.
Among the dangerous possibilities are flat directions of the effective
potential in which combinations such as $L_i Q_3 D_3,~~H_2L_i,~~LLE,~~ H_2L$
acquire vev's. Avoiding these vacua imposes constraints that depend on the
soft supersymmetry breaking parameters: they are weakest for $A\simeq
m_{1/2}$. Figure 7 illustrates some of the resulting constraints in the
$(m_{1/2}, m_0)$ plane, for different values of $\tan\beta$ and signs of
$\mu$ \cite{AF}. We see that they cut out large parts of the plane, particularly for low
$m_0$. In combination with cosmology, they tend to rule out large values of
$m_{1/2}$, but this aspect needs to be considered in conjunction with the
effects of co-annihilation, that are discussed in the next section.

\section*{Co-Annihilation Effects on the Relic Density}
As $m_{\chi}$ increases, the LSP annihilation cross-section decreases and 
hence its relic
number and mass density increase. How heavy could the LSP be?  Until 
recently, the limit
given was $m_{\chi} \lappeq 300$ GeV~\cite{limit}.  However, it has now
been pointed 
out that there are
regions of the MSSM parameter space where co-annihilations of the $\chi$ 
with the stau
slepton $\tilde{\tau}$ could be important, as seen in Fig.8~\cite{EFO}.
These 
co-annihilations would
suppress $\Omega_{\chi}$, allowing a heavier neutralino mass, and we now 
find that~\cite{EFO}
\beq
m_{\chi} \lappeq \; 600 \, \mbox{GeV}
\label{thirtyseven}
\eeq
is possible if we require $\Omega_\chi h^2 \leq 0.3$.  In the past, it was thought that all the 
cosmologically-preferred region of
MSSM parameter space could
be explored by the LHC~\cite{Abdullin}, as seen in Fig. 9, 
but it now seems
possible that there may be a delicate region close to the upper bound 
(\ref{thirtyseven}). 
This point requires further study.

\begin{figure}[htb] 
\centerline{\epsfig{file=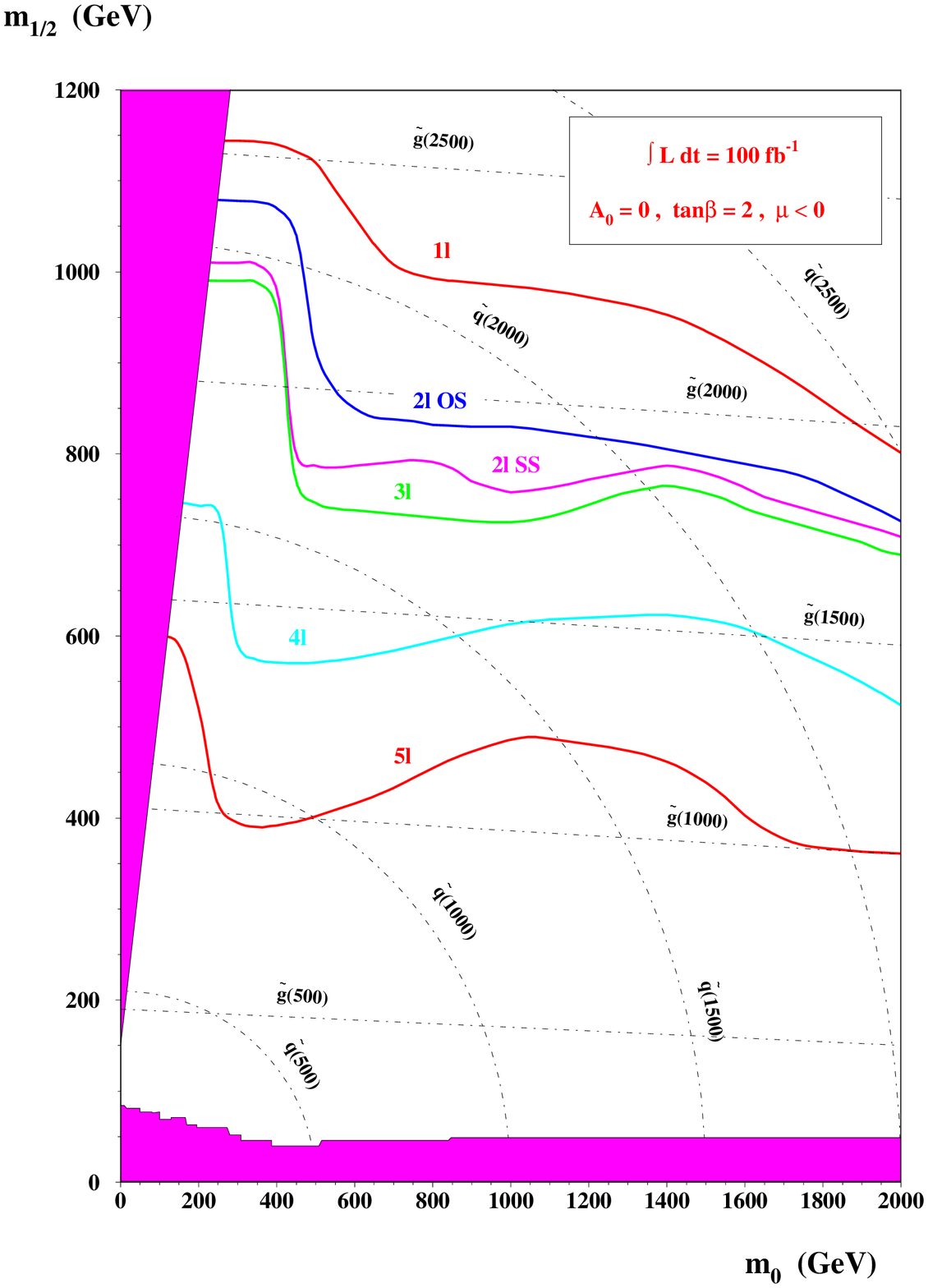,height=3.5in,width=3.5in}}
\vspace{10pt}
\caption{The region of the $(m_0, m_{1/2})$ plane accessible
to sparticle searches at the LHC~[22].}
\label{fig9}
\end{figure}

 \section*{Current LEP Constraints}
 The LEP constraints on MSSM particles have recently been updated \cite{LEPC},
 constraining the parameter space and hence the LSP. The large luminosity
 accumulated during 1998 has enabled the lower limit on the chargino mass to
 be increased essentially to the beam energy: $m_{\chi^\pm} \gappeq$ 95 GeV,
 except in the deep Higgsino region, where the limit decreases to about 90 GeV
 because of the small mass difference between the chargino and the LSP, which reduces the efficiency for
 detecting the $\chi^\pm$ decay products. There are also useful limits on
 associated neutralino production $e^+e^- \rightarrow \chi\chi^\prime$, which
 further constrain the LSP. Without further theoretical assumptions, the
 purely experimental lower limit on the neutralino mass has become
 \beq
 m_\chi \gappeq 32~{\rm GeV}
 \label{one}
 \eeq
 for large values of $m_0$ whatever the value of $\tan\beta$, decreasing to a
 minimum of 28 GeV for small $m_0$.
 
 There are other new LEP limits that come into play with supplementary
 theoretical assumptions. These include a lower limit on the slepton mass,
 assuming universality $(m_{\tilde l}\equiv m_{\tilde e} = m_{\tilde\mu} =
 m_{\tilde\tau})$:
 \beq
 m_{\tilde l} > 90~{\rm GeV}
 \label{two}
 \eeq
 for $m_{\tilde l} - m_\chi \gappeq$ 5 GeV. There is also a new lower limit
 \beq
 m_{\tilde t} > 85~{\rm GeV}
 \label{three}
 \eeq
 assuming the dominance of $\tilde t \rightarrow c\chi$ decay, for $m_{\tilde
 t} - m_\chi \gappeq$ 10 GeV. Most important, however, is the new lower limit
 on the mass of the lightest Higgs boson in the MSSM. The L3 collaboration
 reports
 \beq
 m_h > 95.5~{\rm GeV}
 \label{four}
 \eeq
 for $\tan\beta \lappeq 3$. Combining all four LEP experiments, the lower
 limit (17) would probably be increased to 98 GeV, corresponding to the
 kinematic limit $\sqrt{s}$ (= 189 GeV) - $m_Z$.
 
 The MSSM Higgs and other limits now appear to effectively exclude the
 possibility of Higgsino dark matter. Moreover, for $\mu < 0$, we now find $\tan\beta \gappeq$ 3.0,
 whereas a slightly smaller value is allowable if $\mu > 0$.
 For values of $\tan\beta$ close to these lower limits, the lower limit
 on $m_\chi$ increases sharply, qualitatively as in Fig. 6 but now shifted to
 the right. The valley in Fig. 6a for $\mu < 0$ is now filled in, so, pending
 a more complete evaluation, we estimate that
 \beq
 m_\chi \gappeq 50~{\rm GeV}
 \label{five}
 \eeq
 for either sign of $\mu$.

 \section*{Summary}
 We have seen that current experimental constraints impose $m_\chi \gappeq$ 50
 GeV if universal soft supersymmetry breaking mass parameters are assumed, and
 that $m_\chi \lappeq$ 600 GeV if we require $\Omega_\chi h^2 \leq 0.3$.
 Values of $m_\chi$ close to the lower limit may be explored by forthcoming
 runs of LEP in 1999 and 2000: the searches for the Higgs boson will be
 particularly interesting to follow. Thereafter, Run II of the Tevatron
 collider has the best accelerator chances to find supersymmetry, until the
 LHC comes along.
 
 In the mean time, non-accelerator searches looking directly for LSP-nucleus
 scattering or indirectly at LSP annihilation products will be offering stiff
 competition. There is already one direct search that does not claim not to
 have observed LSP-nucleus scattering \cite{DAMA}. The possible signal would correspond to
 a domain of MSSSM parameter space close to the present limits. The LSP
 interpretation of the signal is not yet generally accepted, since a complete
 annual modulation cycle has not yet been reported. However, healthy
 scepticism should not obscure the fact that it is consistent with the limits
 on sparticle dark matter reported here. Time only will tell whether
 accelerator or non-accelerator experiments will win the race to discover
 supersymmetry.

\end{document}